\documentclass[smallextended]{svjour3}
\smartqed
\usepackage{graphicx}
\usepackage{amsmath}
\usepackage{amssymb}

\begin{document}

	\title{Quantum key distribution over noisy channels by the testing state method}
	\author{Hao Shu, Chang-Yue Zhang, Yue-Qiu Chen, Zhu-Jun Zheng, Shao-Ming Fei}
	\institute{Hao Shu\at
		Shenzhen University
		\\
		South China University of Technology
		\\
		\email{Hao\_B\_Shu@163.com}
		\\
		Chang-Yue Zhang\at
		South China University of Technology
		\\
		Yue-Qiu Chen\at
		Zhongkai University of Agriculture and Engineering
		\\
		South China University of Technology
		\\
		Zhu-Jun Zheng\at
		South China University of Technology
		\\
		Shao-Ming Fei\at
		Capital Normal University
		\\
		Max-Planck-Institute for Mathematics in the Sciences
	}

	\begin{abstract}
		Quantum key distribution(QKD) might be the most famous application of quantum information theory. The idea of QKD is not difficult to understand but in practical implementations, many problems are needed to be solved, for example, the noise of the channels. Previous works usually discuss the estimate of the channels and employ error-correcting procedures, whose feasibility and efficiency depend on the strength of the noise, or assist with entanglement distillation procedures, which often result in a large consumption of states while not all states can be distilled. This paper aims to study QKD over noisy channels including Pauli noises, amplitude damping noises, phase damping noises, collective noises as well as mixtures of them, in any strength without distillations. We provide a method, called the testing state method, to implement QKD protocols without errors over arbitrarily strength noisy channels. The method can be viewed as an error-correcting procedure, and can also be employed for other tasks.
		
		
	\end{abstract}

	\maketitle
	
	\section{Introduction}
	
	 Secure communication is one of the most important subjects in the information era. As Shor's algorithm\cite{S1994Algorithms} can break the security of the most-used RSA public-key system, searching for new cryptography schemes becomes imperative.
	
	 Suppose that the legitimate partner, Alice and Bob, wants to communicate. The most secure method is using a one-time pad as a private key. Hence, the main problem is how to transmit such a key securely. Despite classical communication might not solve this problem, quantum key distribution(QKD) gives a solution. Generally, QKD assumes that Alice and Bob have both insecure quantum channels and authenticated classical channels which might not be private, namely the eavesdropper, Eve, may eavesdrop on classical communications but can not forge or tamper, while she can access quantum channels without any restriction except physical laws. The security criterion is that if Eve gets enough information about the key, then she is detectable by the legitimate partner who can abort the key.
	
	There are mainly two classes of QKD protocols including prepare-measure ones\cite{BB1984Quantum,BB1992Quantum,B1998Optimal,CB2002Security,SP2000Simple,E1991Quantum,TP2018Kak's,K2006A} and entanglement-based ones\cite{SP2000Simple,LC1999Unconditional,ST2016A,GR2010Quantum,ZZ2021Entanglement}. However, no matter which class the protocols belong to, a restriction in implementing is the channel noises\cite{ST2016A,ST2015Which,S2016Effect,SB2018Analysis}. The most researched noises might be collective noises such as collective dephasing (CD) and collective rotation (CR) \cite{LZ2009Fault,LD2008Efficient,SD2010Efficient,BG2004Robust}. Other researches include Pauli noises\cite{SS2007Degenerate,FW2008Lower,FM2001Enhanced,CR2011Experimental}, amplitude damping (AD) and phase damping (PD) noises\cite{TP2015Applications,SS2015Controlled,OS2013Dissipative,TM2000Decoherence,KL2012Protecting,XY2016Protecting}, as well as the related ones\cite{OS2013Dissipative,SB2008Squeezed,SO2012The,TB2015Quasiprobability,TB2016Tomograms}.
	
	Previous works on noises mainly include noise estimate and error-correcting procedures, whose feasibility and efficiency depend on the strength of the noises, as well as entanglement distillation procedures, which often result in a large consumption of states while not all states can be distilled\cite{BB1996Purification,CL1996How,F2014PPT,PM2007Class,BR2003Classes,VC2001Irreversibility,EV2001Distillability,BD1996Mixed,D2016On}.
	
	In this paper, we study QKD schemes with Pauli noises, AD and PD noises, collective (CP and CR) noises, as well as their mixtures. We give a method against such noises with any strength by using testing states, in which entanglement distillations are not needed. Our scheme can be implemented in noisy channels without errors, like in noiseless ones.
	
	\section{General description of the scheme}
	
	In the rest of the paper, we assume that if two states are sent together via the same channel, then they will suffer the same effects and since we only focus on the channels, other devices are assumed to be idealized. We will call the presenting scheme \textit{testing state method}, which is employed to modify QKD schemes, making them robust over noises. In the scheme, the sender sends key states and testing states together such that they encounter the same effects. The key states are employed for the key as in the ordinary protocol while the testing states are employed to modify the bits in the key, See the figure.
	
	\vspace{0.5cm}
	\includegraphics[width=1.0\textwidth]{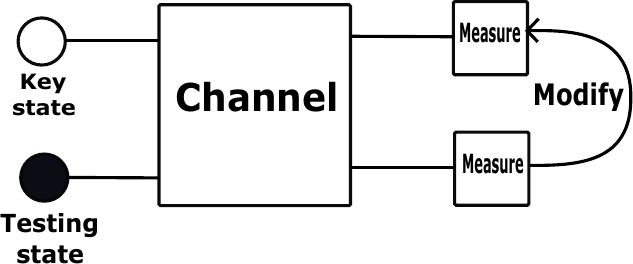}

    \vspace{0.5cm}
    The general description which detailed the figure is as follows with employments for different kinds of noises in the next section.
 \\

	\textbf{Step 1:} The sender prepares key states and testing states. The key states are used to generate the key and estimate security as usual, while the testing states are used to modify the bits over noisy channels.
	
	\textbf{Step 2:} The sender sends the states via channels, where the key states and the testing states are sent together and assumed to suffer the same effects.
	
	\textbf{Step 3:} The legitimate partner measures each testing state via fixed bases and measures each key state as in the ordinary protocol. The outcomes(bits) of key states should be modified by the outcomes of the testing states.
	
	\textbf{Step 4:} The legitimate partner calculates the error rate and generates a raw key by key states, then runs error-correcting and privacy amplification procedures if needed.
	
	In the above scheme, the choice of testing states depends on the channel, namely what kind of noises the sending states might encounter. In the rest of the discussions, we take BBM92 protocol\cite{BB1992Quantum} as the ordinary protocol to illustrate how the testing state method works but note that the method is also suitable for other protocols such as BB84 protocol.
	
	The standard BBM92 protocol includes Alice prepares maximally entangled states $|\varphi \rangle_{AB}=\frac{1}{\sqrt{2}}\sum_{j=0}^{1}|j\rangle_{A}|j\rangle_{B}$, where $\left \{ |j\rangle |\,j=0,1\right \}$ is an orthonormal basis in $C^2$, sends the B partita to Bob via quantum channels and they measure each state via randomly chosen basis $\left \{ |0\rangle, |1\rangle\right \}$ or $\left \{ |+\rangle=\frac{1}{\sqrt{2}}(|0\rangle+ |1\rangle),\,|-\rangle=\frac{1}{\sqrt{2}}(|0\rangle- |1\rangle)\right \}$, as well as basis sifting, error estimating and key generating procedures.
	
	\section{Testing state method in different noisy channels}
	
	 In the following, we investigate how BBM92 protocol can resist certain noises with the scheme. Note that the discussions in the following can be directly extended to other protocols. By letting Alice send key states and testing states together, we assume that they will encounter the same effects. In the subsections below, we will discuss how to choose testing states for special noises.
	
		\subsection{Pauli noises}
		
		Pauli noises act on each qubit sent by the channel via Pauli operators,
		I=$\begin{bmatrix}
			1  & 0\\
			0  & 1
		\end{bmatrix}$,
		Z=$\begin{bmatrix}
			1  & 0\\
			0  & -1
		\end{bmatrix}$,
		X=$\begin{bmatrix}
			0  & 1\\
			1  & 0
		\end{bmatrix}$,
		ZX=$\begin{bmatrix}
			0  & 1\\
			-1  & 0
		\end{bmatrix}$ under the computational basis, with corresponding probabilities $p_{I}$, $p_{Z}$, $p_{X}$ and $p_{ZX}$, summing to 1. Note that it is not the scenario that the states suffer general noises, which can be described as phase errors and bit errors, but the scenario that the channel can be described by the Kraus operators $\{\sqrt{p_{I}}I, \sqrt{p_{X}}X, \sqrt{p_{Z}}Z, \sqrt{p_{ZX}}ZX\}$.
		
		\subsubsection{Two-Pauli channel}
		
		In two-Pauli channels, $p_{I}$, $p_{Z}$, $p_{X}$, $p_{ZX}$ might all be non-zero. The final states might be completely mixed. The modified protocol is described as follows.
		\\
		
		 \textbf{Step 1:} To share a N-bit key string, Alice prepares 2N states $|\varphi\rangle_{AB_{1}}|0\rangle_{B_{2}}|+\rangle_{B_{3}} = \frac{1}{\sqrt{2}}(|00\rangle+|11\rangle)_{AB_{1}}|0\rangle_{B_{2}}|+\rangle_{B_{3}}$. Here, partite $A, B_{1}$ are for keys while partite $B_{2}, B_{3}$ are for testing.
		
	    \textbf{step 2:} Alice sends B partite (including $B_{1}, B_{2}, B_{3}$) to Bob such that the B partite of a state are always sent together and assumed to suffer the same effects.
		
		\textbf{step 3:} For each state, Bob measures partite $B_{2}$ and $B_{3}$ via basis $\left \{ |0\rangle, |1\rangle\right \}$ and $\left \{ |+\rangle, |-\rangle\right \}$, respectively, while Alice and Bob measure partite $A, B_{1}$ randomly via basis $\left \{ |0\rangle, |1\rangle\right \}$ or $\left \{ |+\rangle, |-\rangle\right \}$. The testing bits are employed for testing which operator affected the key state and the key bits are modified by the testing bits.
		
	    \textbf{Step 4:} After the basis sifting procedure, Alice decides her bits by outcomes on partita A as usual while Bob decides his bits by outcomes on partite $B_{1}, B_{2}, B_{3}$. Precisely, Bob decides a bit by the outcome on partita $B_{1}$ as usual but flips it if the outcome on $B_{2}$ is 1 when he chose basis $\left \{ |0\rangle, |1\rangle\right \}$ to measure the key state, or if the outcome on $B_{3}$ is -, when he chose basis $\left \{ |+\rangle, |-\rangle\right \}$ to measure the key state.
		\\
		
		Other procedures are the same as in the ordinary protocol. In such a protocol, all key bits can be employed for the key.
		
		\subsubsection{One-Pauli channel}
		
		In one-Pauli channels, states suffer two of the four Pauli operators, one of which is I.
		
		Let us assume that states suffer I with probability p and Z with probability 1-p. The procedures are exactly the same as above. However, partita $B_{2}$ can be saved, namely, Alice and Bob use $|\varphi\rangle_{AB_{1}}|+\rangle_{B_{3}}$ instead. This is because they only need to test whether the states are affected by Z, which can be accomplished by partita $B_{3}$ only.
		
		Other kinds of one-Pauli channels can be analyzed exactly in the same way. If the states suffer I with probability p and X with probability 1-p, $B_{3}$ can be saved. And if the states suffer I with probability p and ZX with probability 1-p, either $B_{2}$ or $B_{3}$ can be saved.
		
		Similar to the two-Pauli channel, all key bits can be employed for the key.
		
	\subsection{Phase damping (PD)}
	
	Phase damping noises have kraus operators
	$E_{0}=\begin{bmatrix}
		\sqrt{1-p}  & 0\\
		0           & \sqrt{1-p}
	\end{bmatrix}$,
	$E_{1}=\begin{bmatrix}
		\sqrt{p}  & 0\\
		0  & 0
	\end{bmatrix}$,
	$E_{2}=\begin{bmatrix}
		0  & 0\\
		0  & \sqrt{p}
	\end{bmatrix}$.
	Hence, a state sent through the channel remains unchanged with probability 1-p and suffers an error with probability p.

	In this case, Alice prepares 2N states $|\varphi\rangle_{AB_{1}}|0\rangle_{B_{2}}|1\rangle_{B_{3}} = \frac{1}{\sqrt{2}}(|00\rangle+|11\rangle)_{AB_{1}}|0\rangle_{B_{2}}|1\rangle_{B_{3}}$ instead, where partite $B_{2}$ and $B_{3}$ are for testing and others are for keys (Step 1). Alice always sends the three B partite of a state together and they are assumed to suffer the same effect (Step 2). On average, only 1-p of the states can be received without any loss, which are not affected by noises. For such states, Alice and Bob run other procedures as in the ordinary protocol (Steps 3, 4). As a consequence, 1-p of the key states can be employed for the key.

\subsection{Amplitude damping (AD)}

Amplitude damping noises are given by Kraus operators
$E_{0}=\begin{bmatrix}
	1  & 0\\
	0  & \sqrt{1-p}
\end{bmatrix}$ and
$E_{1}=\begin{bmatrix}
	0  & \sqrt{p}\\
	0  & 0
\end{bmatrix}$. Hence, a state sent through the channel has probability 1-p suffering $E_{0}$ and probability p suffering $E_{1}$.

To deal with such a noise. Alice prepares 2N states $|\varphi_{p}\rangle_{AB_{1}}|0\rangle_{B_{2}} = \frac{1}{\sqrt{2-p}}(\sqrt{1-p}|00\rangle+|11\rangle)_{AB_{1}}|0\rangle_{B_{2}}$ instead, where partite $A, B_{1}$ are for keys, partita $B_{2}$ is for testing (Step 1). Alice always sends partite B of each state together such that they suffer the same effects. On average, only 1-p of the states can be received without any loss, which become $|\varphi\rangle_{AB_{1}}|0\rangle_{B_{2}} = \frac{1}{\sqrt{2}}(|00\rangle+|11\rangle)_{AB_{1}}|0\rangle_{B_{2}}$ (Step 2). For such states, Alice and Bob run other procedures as usual, ignoring testing states (Steps 3, 4). Similar to the PD channel, 1-p of the key states can be employed for the key.

    \subsection{Collective noises}

    To deal with collective noises, we follow the method called decoherence-free states and previous works\cite{LZ2009Fault,LD2008Efficient,SD2010Efficient,BG2004Robust}.

	\subsubsection{Collective dephasing (CD)}
	
	The collective dephasing noises act on each qubit sent through the channel equivalently via
	$\begin{bmatrix}
		1  & 0\\
		0  & e^{i\phi}
	\end{bmatrix}$
	under the computational basis, where $\phi$ is a parameter depending on the noise.
	
	Note that $|\psi'\rangle_{B_{1}B_{2}} = \frac{1}{\sqrt{2}}(|01\rangle-|10\rangle)_{B_{1}B_{2}} = \frac{1}{\sqrt{2}}(|-+\rangle-|+-\rangle)_{B_{1}B_{2}}$ and $|\psi\rangle_{B_{1}B_{2}}= \frac{1}{\sqrt{2}}(|01\rangle+|10\rangle)_{B_{1}B_{2}} = \frac{1}{\sqrt{2}}(|++\rangle-|--\rangle)_{B_{1}B_{2}}$ change nothing but a global phase under such a noise. Therefore, let Alice uses state $\frac{1}{\sqrt{2}}(|0\rangle|\psi'\rangle+|1\rangle|\psi\rangle)_{AB_{1}B_{2}}$ instead, sending partite $B_{1}, B_{2}$ to Bob together. After Bob receives a state, he operates a unitary transformation, transforming $|\psi'\rangle_{B_{1}B_{2}}, |\psi\rangle_{B_{1}B_{2}}$ to $|00\rangle_{B_{1}B_{2}}, |10\rangle_{B_{1}B_{2}}$, respectively, and aborts partita $B_{2}$. Hence, they obtain $|\varphi\rangle_{AB_{1}}$ and other procedures are as usual. Hence, all key bits can be employed for the key.
	
	\subsubsection{Collective rotation (CR)}
	
	Collective rotation noises act on each qubit sent by the channel equivalently via
	$\begin{bmatrix}
		cos\theta  & sin\theta\\
		sin\theta  & -cos\theta
	\end{bmatrix}$
	under the computational basis, where $\theta$ is a parameter depending on the noise and evolves upon time.
	
	Both $|\varphi\rangle_{B_{1}B_{2}}=\frac{1}{\sqrt{2}}(|00\rangle+|11\rangle)_{B_{1}B_{2}}$ and $|\psi'\rangle_{B_{1}B_{2}}=\frac{1}{\sqrt{2}}(|01\rangle-|10\rangle)_{B_{1}B_{2}}$ are unchanged upon a global phase. Therefore, Alice uses states $\frac{1}{\sqrt{2}}(|0\rangle|\varphi\rangle+|1\rangle|\psi'\rangle)_{AB_{1}B_{2}}$ instead. Others are similar to the CD case. Hence, similar to the CD case, all key bits can be employed for the key.
	
	\subsection{Mixture of different noises }
	
	We can deal with mixtures of different noises by combining the above strategies. Let us discuss two kinds of them.
	
	\subsubsection{Mixture of PD, Pauli noises, and CD}
	
	It is enough to use state $\frac{1}{\sqrt{2}}(|0\rangle|\psi'\rangle+|1\rangle|\psi\rangle)_{AB_{1}B_{2}}$. Alice sends the B partite together such that the state is immune from Pauli and CD noises. If it is affected by PD noises, it will be lost. Therefore, if Bob receives without any loss, then the state is not affected by such noises. Hence, all key bits can be employed for the key.
	
	\subsubsection{Mixture of PD, Pauli noises, and CR}
	
	Similar to above, Alice uses state $\frac{1}{\sqrt{2}}(|0\rangle|\varphi\rangle+|1\rangle|\psi'\rangle)_{AB_{1}B_{2}}|\psi'\rangle_{B_{3}B_{4}}$ instead. Here, if Bob receives B partite without any loss, the state is not affected by PD noises. Then the legitimate partner uses partite $A, B_{1}, B_{2}$ to generate a key since they are immune from CR and Pauli noises. Similar to the last subsection, all key bits can be employed for the key.
	
	\section{Discussion}
	
	The security of the protocols comes from the security of the ordinary protocols since testing states are not private and are only used to correct errors. As the testing states are independent of the encoding states and carry no private information, they provide no benefits to the eavesdropper.
	
	The arguments can be viewed as an error-correcting procedure and can be employed for other tasks or protocols\cite{S2021Quantum}, such as BB84 protocol and protocols with an untrusted third party who distributes entangled states to Alice and Bob. Although BBM92 protocol might not be a mostly employed one, we choose it as an example because it is easy to understand and can demonstrate that the scheme does not need to use entanglement distillations.
	
	The state $|\psi'\rangle$ is unchanged under collective noises. Such a state is called a decoherence-free state. For a mixture of PD, CP, CR, and Pauli noises, in general, we can use two orthogonal decoherence-free ones to handle. However, this is impossible in $C^{2}\otimes C^{2}$, since $|\psi'\rangle$ is the unique state with this property. But in $C^{4}\otimes C^{4}$, this can be done \cite{C2007Six}.
	
    Note that the above method requires knowing the kinds of noise over the channel and thus estimating the channel should come in the first place. Although it is beyond this paper, we mention that this can be done by other technologies such as state tomography\cite{S2021Informationally}.
    
    The previous sections indicate that the method should work better in the setting of Pauli noises since no entangled states are needed for testing, which implies that only separated states are required when applying to protocols without entanglement such as the BB84 protocol, while worse in cases of collective noises because three states entanglement is needed, which might difficult to implement in practice (Note that the efficiency of two state entanglement might have been insufferable).
    
    We do not particularly mention the experimental settings here since the method is a theoretical scheme for general protocols. The settings of experiments depend on what realizations are employed practically. For example, usually, optical devices can be employed.
	
	Finally, we mention that noise processes are not the only problems in communication tasks. Other problems might include, for example, signal improved\cite{SB2020Quantum}, detectors\cite{SB2022Analysis,S2022Solving}, and sources\cite{TS2019Coherent}.
	
	\section{Conclusion}
	
	In this paper, we have investigated a method for quantum key distribution over certain noisy channels and presented modified protocols, which can be implemented over noisy channels with any strength like in noiseless ones, without giving rise to errors. The method is using testing states to correct bits and is thus called the testing state method. We took the BBM92 protocol as an example to illustrate how it works but it can also be extended to other protocols, including QKD ones and others, directly.
	
	\section*{Acknowledgement}
	
	This work is supported by the Key Research and Development Project of Guangdong province under Grant No.2020B0303300001, the Guangdong Basic and Applied Research Foundation under Grant No.2020B1515310016.
	
	\section*{Data availability}
	
	The author declare that all data supporting the findings of this study are available within the paper.
	
	\section*{Competing interests}
	
	\qquad The author declare no competing interests.
	
	\bibliographystyle{unsrt}
	\bibliography{Bibliog}
\end{document}